\documentclass[11pt]{article}
\usepackage[centertags]{amsmath}
\usepackage{amssymb}
\usepackage[dvips]{graphicx}
\usepackage{epsfig,graphics}

\def\vF{{\bf F}}

\def\vx{{\bf x}}
\def\vJ{{\bf J}}

\begin{document}
\renewcommand{\baselinestretch}{1.5}

\title{\bf Landscapes of Non-gradient Dynamics Without
Detailed Balance: Stable Limit Cycles and Multiple
Attractors}

\author{Hao Ge$^{1}$\footnote{Email: haoge@pku.edu.cn}
and
Hong Qian$^{2}$\footnote{Email: hqian@u.washington.edu}
\\
$^1$Beijing International Center for Mathematical Research (BICMR) \\
and Biodynamic Optical Imaging Center (BIOPIC)\\
Peking University, Beijing, 100871, PRC\\
$^2$Department of Applied Mathematics\\
University of Washington, Seattle, WA 98195, USA
}

\maketitle{}

\newpage
\pagebreak

\renewcommand{\baselinestretch}{1}

\begin{abstract}
Landscape is one of the key notions in literature
on biological processes and physics of complex
systems with both deterministic and stochastic dynamics.
The large deviation theory (LDT) provides a possible
mathematical basis for the scientists' intuition.
In terms of Freidlin-Wentzell's LDT, we discuss
explicitly two issues in singularly perturbed
stationary diffusion processes arisen from nonlinear
differential equations:
(1) For a process whose corresponding ordinary
differential equation has a stable
limit cycle, the stationary solution exhibits a clear
separation of time scales: an exponential terms
and an algebraic prefactor. The large deviation
rate function attains its minimum zero on the entire
stable limit cycle, while the leading term of the prefactor is
inversely proportional to the velocity of the non-uniform
periodic oscillation on the cycle. (2) For dynamics with
multiple stable fixed points and saddles, there is in
general a breakdown of detailed balance among the
corresponding attractors.  Two landscapes, a local and a
global, arise in LDT, and a Markov jumping process with
cycle flux emerges in the low-noise limit. A local
landscape is pertinent to the transition rates between
neighboring stable fixed points; and the global
landscape defines a nonequilibrium steady state.
There would be nondifferentiable points in the latter for a
stationary dynamics with cycle flux. LDT serving as the
mathematical foundation for emergent
landscapes deserves further investigations.
\end{abstract}


\newpage
{\bf
Stochastic nonlinear approaches to dynamics has attracted
great interests from physicists, biologists, and
mathematicians in current research.  More than 70 years ago,
Kramers has developed a diffusion model characterizing
the molecular dynamics along a reaction coordinates,
via a barrier crossing mechanism, and calculated the
reaction rate for an emergent chemical reaction.
The work explained the celebrated Arrhenius relation as well
as Eyring's concept of ``transition state''. Kramers' theory, however, is only valid for stochastic dynamics
in closed systems with detailed balance (i.e., a gradient
flow), where the energy landscape gives the equilibrium
stationary distribution via Boltzmann's law.
It is not suitable for models of open systems.  Limit cycle
oscillation is one of the most important emergent behaviors
of nonlinear, non-gradient systems.  The large deviation
theory from probability naturally provides a basis for
the concept of a ``landscape'' in a deterministic nonlinear, non-gradient dynamics.  In the present study, we initiate a line of
studies on the dynamics of and emergent landscape in
open systems.  In particular, using singularly perturbed
diffusion on a circle as a model, we study systems
with stable limit cycle as well as systems with multiple
attractors with nonzero flux. A seeming paradox
concerning emergent landscape for limit cycle is
resolved; a local theory for transitions between
two adjacent attractors, \`{a} la Kramers, is discussed;
and a ``$\lambda$-surgery'' to
obtain nonequilibrium steady state (NESS) landscape
for multiple attractors is described.
}

\section{Introduction}

    Stochastic nonlinear dynamics (SND) of biochemical reaction
systems at the cellular and subcellular level has received
much interests in recent years from applied mathematicians, physicists, as well as biologists \cite{qian_iop}.  In
terms of stochastic processes, the main mathematical
approaches to biochemical SND are either diffusion processes
i.e., the chemical Langevin equation, or
the chemical master equation \cite{wilkinson,qian_book,gillespie}
which characterizes
the evolution of probability distribution for a Markov
jump process that can be simulated by the method of
Gillespie algorithm.  For both stochastic models, their
infinite large system (macroscopic) limit is a system of
nonlinear ordinary differential equations (ODE) based
on the Law of Mass Action \cite{kurtz,kurtz_book}.
These mathematical models have provided a unique opportunity for
comparative studies of corresponding nonlinear dynamics in
small and in large systems.

\begin{figure}[t]
\begin{center}
   \includegraphics[width=5.25in]{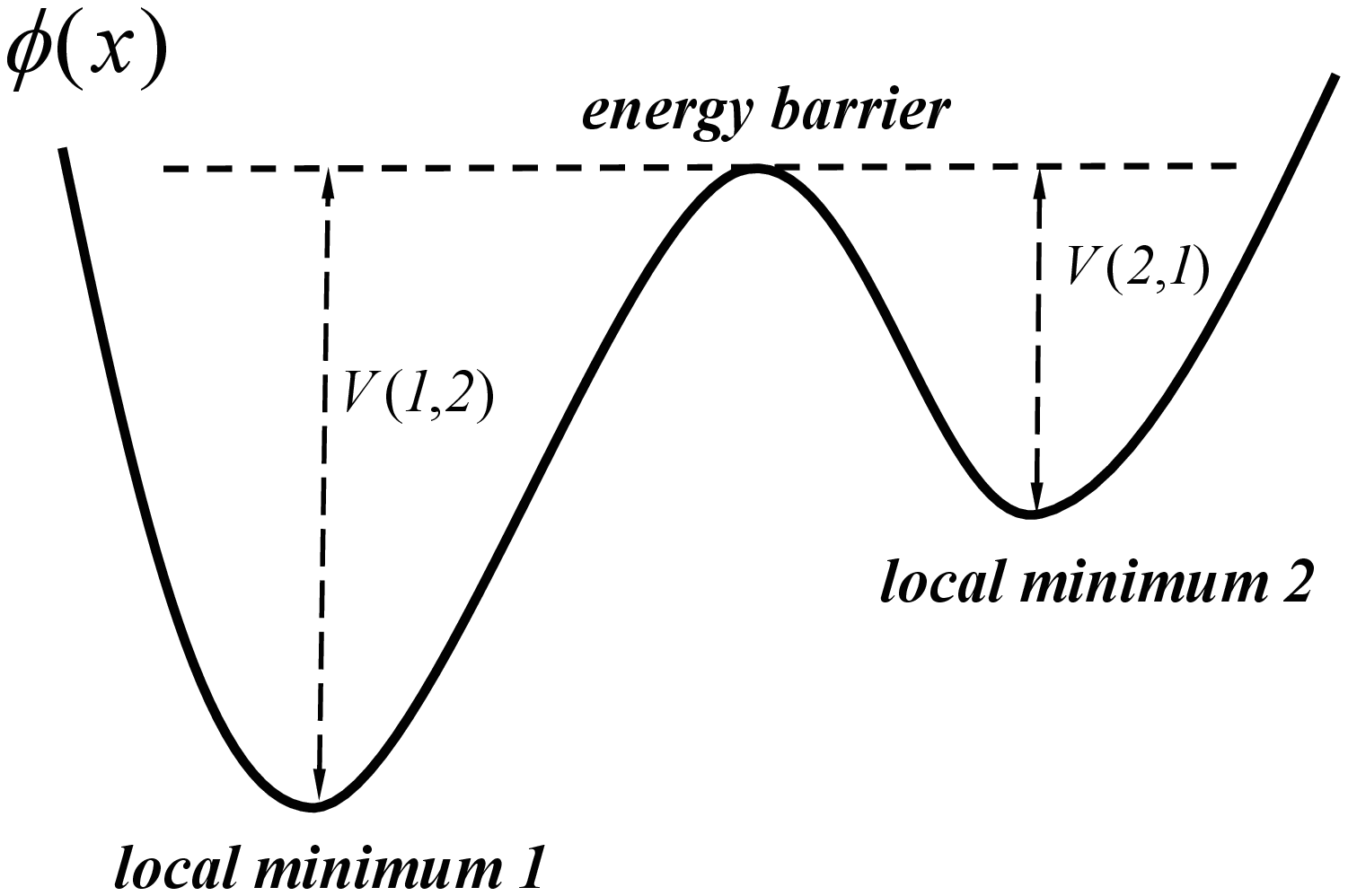}
   \caption{Landscape $\phi(x)$ and related Kramers'
rate theory for a bistable system.  The local minima
correspond to {\em stable} fixed points of a deterministic
dynamics $\dot{x}=-d\phi(x)/dx$ while the maximum
corresponds to an {\em unstable} fixed (saddle) point.
The $V(1,2)$ and $V(2,1)$ represent
the energy barriers for exiting energy wells $1$ and $2$, respectively.  For very small $\epsilon$, the stationary
probability distribution for stochastic dynamics with Brownian
motion $B(t)$, $dx=-(d\phi(x)/dx)dt+\sqrt{2\epsilon}\ dB(t)$,
is $u_{\epsilon}(x) \propto \exp\big(-\phi(x)/\epsilon\big)$.
The Kramers theory yields transition rates between the two attractors:
$k_{12}\propto  e^{-V(1,2)/\epsilon}$ and
$k_{21}\propto e^{-V(2,1)/\epsilon}$.  According to Freidlin-Wentzell's LDT, this theory still applies for every pair of
neighbouring attractors of a non-gradient system in terms of
a {\em local landscape}. However, the stationary probability
distribution follows a different, {\em global landscape}.  Also
see Fig. \ref{fig3}.}
  \label{chaos_fig1}
  \end{center}
\end{figure}

	In quantitative biology and in statistical physics, there
is an emerging notion of ``landscape'' for dynamics
\cite{wolynes_1}, as both a metaphor and as an analytical
device \cite{wolynes_2,ao_2006,gehao_2}.
Landscape for a gradient system is of course natural, and the well-known Kramers' rate theory directly follows \cite{Kramers} (Fig. \ref{chaos_fig1}). For a non-gradient system, a
Lyapunov function \cite{perko}, if exists, can still be visualized as a landscape for the dynamics.  The real question is whether
there always exists such a landscape function for non-gradient system and whether there is a corresponding Kramers'-like rate formula for the inter-attractoral transition rates.  For
bistable systems, the answer to this question is yes  \cite{freidlin,gehao_1,gehao_2,gehao_mcb}: The rate formula
is exactly the same as the classical Kramers' formula
(Fig. \ref{chaos_fig1}).  But how could one
perceive a ``landscape'' for a periodically oscillatory
system?  Furthermore, is there any difference between
multistable and bistable systems?  These questions
are the motivations of the present paper.

The case of a system with a limit cycle
is in defiance of the intuition \cite{ao_2006}, and there are other serious, but subtle arguments
against the general notion of landscape for systems with limit
cycles.  Noting that a same landscape is used
for a deterministic dynamics as well as the stationary
probability of its stochastic counterpart,
one objection can be stated as follows: Let
$\phi(x)$ be a landscape of a system with limit cycle
$\Gamma$.  Then $\phi(x)$ has to be a constant on $\Gamma$.
However, since the landscape is also expected to represent
the probability of a stochastic system: Lower
$\phi$ corresponds to higher probability.  Combining the
two lines of reasoning, one arrives at equal probability
along $\Gamma$.  Now according to the ergodic theory,
equal probability on $\Gamma$ implies uniform velocity
on the limit cycle.  This suggests that only uniform
rotation is compatible with the notion of a landscape
\cite{ao_2006}.

	One of the aims of the present paper is to
give an explicit resolution to this seeming
paradox. The analysis reveals a separation of time
scales for stochastic dynamics and its deterministic
limit.  Indeed around a limit cycle $\Gamma$, the probability
$u_{\epsilon}(x) \simeq C_0(x)e^{-\phi(x)/\epsilon}$
in which $\phi(x)=0$ along the $\Gamma$. The dynamics on the limiting set $\Gamma$, therefore, is determined by $C_0(x)$.

For presenting
the results, we choose to be insightful rather than
thorough and rigorous.
Hence we shall only discuss the problem in terms of
singularly perturbed diffusion processes.  The insights
we obtain, however, are qualitatively applicable also
to other systems, even though technically they might be
much more difficult to handle. More precisely, we would like to carry out an analysis of the singularly perturbed
stationary diffusion equation in the form
\begin{equation}
     \nabla\cdot\left(\epsilon\nabla u_{\epsilon}(\vx)
        - u_{\epsilon}(\vx)\vF(\vx)\right) = 0, \ \ \
    (\vx\in\mathbb{R}^N).
\label{s_fpe}
\end{equation}
In Eq. (\ref{s_fpe}) $\epsilon$ is a small positive parameter.

For
a discussion of limit cycles in the Chemical Master Equation, see
\cite{qian_pnas_02,Vellela_3}, and the diffusion approximation in
general, see \cite{tan}. The singularly perturbed $2$nd order linear
elliptic equation is a well-studied problem in mathematics. However,
there are still several important issues remaining unclear, even for
the one-dimensional circle $\mathbb{S}^1$. Here we wish to further
explicitly illustrate some of them in connection to the case
of a stable limit cycle or multiple attractors without detailed balance, sometime using examples. There would be two kinds of landscapes in the case of multiple attractors: one is for the Kramers'-like rate formula and the other is for the global stationary distribution. In a broad sense, both problems are very closely related to
statistical dynamics and thermodynamics of nonequilibrium steady state \cite{zqq,gqq}.

\section{General stationary solution and WKB approximation}

     We assume the function $u_{\epsilon}(\vx)$ in Eq. (\ref{s_fpe}) to be
$L_1$ integrable throughout $\mathbb{R}^N$. And we further
assume that $\vF(\vx)$ is sufficiently well behaved and that a stationary
probability density exists.  See \cite{qian_jsp_02} for appropriate
conditions. It is understood that for Eq. (\ref{s_fpe}), in addition
to the stationary probability $u_{\epsilon}(\vx)$, the system also possesses a
non-trivial flux vector $\vJ$:
\begin{equation}
    \vJ = u_{\epsilon}(\vx)\vF(\vx) -
            \epsilon\nabla u_{\epsilon}(\vx),
\label{ssJ}
\end{equation}
satisfying $\nabla\cdot\vJ=0$.   It can be shown that $\vJ=0$ if and
only if $\vF=-\nabla U$ is a gradient system \cite{qian_jsp_02},
which is called detailed balance or equilibrium. In that case, there
will be no limit cycle and in fact $u_{\epsilon}(\vx)=A\exp(-U(\vx)/\epsilon)$,
where $A$ is a normalization constant.

    Now in terms of the small parameter
$\epsilon$, let us first assume that the limit
\begin{equation}
    \lim_{\epsilon\to 0} \epsilon \ln
        u_{\epsilon}(\vx) = -\phi(\vx) \le 0
\end{equation}
exists.  Note that the limit $\phi(\vx)$ has to
be zero on the entire set where $u_{\epsilon}(\vx)$ has a
nontrivial limit.
In probability theory, $\phi(x)$ is known as the large
deviation rate function \cite{freidlin,qian_1996,roy_93}.
Furthermore, we also assume that the solution has the general form
\begin{equation}
    u_{\epsilon}(\vx) = C_{\epsilon}(\vx) e^{-\phi(\vx)/\epsilon},
\end{equation}
and that there exists a positive constant $\nu$ such that
\begin{equation}
    \lim_{\epsilon\to 0}  \epsilon^{\nu} C_{\epsilon}(\vx)
        = C_0(\vx), \ \ \
        0 < C_0(\vx) < +\infty.
\end{equation}
Therefore, we have the fundamental asymptotic
representation
\begin{equation}
    u_{\epsilon}(\vx) = \epsilon^{-\nu}\left(C_0(\vx)+\epsilon C_1(\vx)
        + \cdots \right)
        e^{-\phi(\vx)/\epsilon},
\label{u_exp}
\end{equation}
as in WKB theory \cite{bender,grasman}.
Substituting Eq. (\ref{u_exp}) into Eq. (\ref{s_fpe}),
we formally have
\begin{eqnarray}
    && \frac{1}{\epsilon}C_0\left(\nabla\phi+\vF\right)
        \cdot\nabla\phi
\nonumber\\[10pt]
    &-& \left(C_0\nabla^2\phi+2\nabla C_0
        \cdot\nabla\phi+\nabla C_0\cdot\vF
        +C_0\nabla\cdot\vF-C_1(\nabla\phi)^2-C_1\vF\cdot\nabla\phi
            \right)
\nonumber\\[10pt]
    &+&  \epsilon\left( \nabla^2 C_0 - C_1\nabla^2\phi-2\nabla C_1
        \cdot\nabla\phi-\nabla C_1\cdot\vF
        -C_1\nabla\cdot\vF\right) + \cdots
            = 0.
\label{eq7}
\end{eqnarray}

The leading order term in (\ref{eq7}) yields
\begin{equation}
    C_0\left(\nabla\phi+\vF\right)\cdot\nabla\phi = 0.
\label{eq8}
\end{equation}
Since $C_0\neq 0$, this means
\begin{equation}
    \vF\cdot\nabla\phi = -\left(\nabla\phi\right)^2 \le 0.
\label{eq9}
\end{equation}
Eq. (\ref{eq9}) shows that for the ordinary differential equation
\begin{equation}
    \frac{d\vx}{dt} = \vF(\vx),
\end{equation}
the function $\phi(\vx)$ has the Lyapunov property:
\begin{equation}
    \frac{d\phi(\vx(t))}{dt}
        = \nabla\phi\cdot\frac{d\vx}{dt}
            =  \nabla\phi\cdot\vF \le 0.
\end{equation}
This result was contained in \cite{matkowsky_1,matkowsky_2,roy_93}
and first explicitly reported in \cite{HuGang} for chemical master
equation.

    The second-order term in Eq. (\ref{eq7}) yields
\begin{equation}
    \nabla C_0\cdot\left( 2\nabla\phi+\vF\right)
        +C_0\left(\nabla^2\phi+\nabla\cdot\vF\right) = 0,
\label{eq12}
\end{equation}
from which $C_0(\vx)$ can be obtained. For example, if
$C_0(\vx)$ is a constant, independent of $\vx$, then Eq.
(\ref{eq12}) implies that $\nabla\phi+\vF=\gamma$ is a
divergence-free vector field.  Combining this with Eq. (\ref{eq8}),
we have
\begin{equation}
    \vF = -\nabla\phi +\gamma, \ \ \
    \nabla\phi\cdot\gamma = 0.
\end{equation}
The vector field $\vF$ thereby has an orthogonal Hodge decomposition
\cite{ludwig_75}.  In other words, if one assumes that
the solution to Eq. (\ref{s_fpe}) is in the form of $e^{-w(\vx)/\epsilon}$,
then $w(\vx)$ has to be a function of $\epsilon$ except when
$\vF$ has orthogonal Hodge decomposition (This is indeed the case for Boltzmann's law with $\gamma=0$).
The leading order expansion is also the starting point of
several investigations carried out by Graham and coworkers
\cite{graham}.  By requiring an orthogonality condition
between the gradient and the rotational parts of the decomposition
of $\vF(\vx)$, the existence of a smooth $\phi(\vx)$ is related
to the complete integrability of certain
Hamiltonian system.   The Lyapunov property of $\phi(\vx)$
in Eq. (\ref{eq9}), however, is more general.

\section{The theory of diffusion on a circle}

\label{1dcircle}

    We now give a thorough treatment of the dynamics on
the circle, which includes either a limit cycle or multiple
fixed points.  We consider the
singularly perturbed, stationary diffusion equation on the circle
\begin{equation}
    \epsilon\frac{d^2u}{d\theta^2}
        +\frac{d}{d\theta}\left\{(U'(\theta)-f)u
            \right\} = 0,
\label{1d_s_fpe}
\end{equation}
in which $U(\theta)$ is a given smooth periodic function,
$U(0)=U(1)$, with periodic boundary condition $u(0)=u(1)$
\cite{omalley_08}, and $f$ is a given constant. The general solution
is
\begin{equation}
    u(\theta) = A_{\epsilon}\left(
        \int_\theta^{1+\theta} e^{\frac{U(z)-fz}{\epsilon}}dz
            \right) e^{-\frac{U(\theta)-f\theta}{\epsilon}},
\label{uss}
\end{equation}
in which $A_{\epsilon}$ is a normalization factor.  An important
quantity associated with the stationary process is the cycle flux
\begin{equation}
    J = \epsilon A_{\epsilon}\left(1-e^{-f/\epsilon} \right),
\end{equation}
which generalizes the rotation number for nonlinear dynamical
systems on the circle \cite{wang_qian}.   When $f=0$, the flux
$J=0$. This is the case of symmetric diffusion process in the theory
of probability \cite{qian_jsp_02}.

We note that in the limit of $\epsilon\rightarrow 0$, by Laplace's
method of integration \cite{bender}, we have
\begin{equation}
    \int_{\theta}^{1+\theta} e^{\frac{U(z)-fz}{\epsilon}}dz
    =  C(\theta,f)\epsilon^{\nu}e^{U^*(\theta)/\epsilon},
\end{equation}
where
\begin{equation}
    U^*(\theta) = \sup_{\theta\le z<1+\theta} \left\{U(z)-fz\right\},
\end{equation}
and $C$ is bounded.  The parameter $\nu$ is either $\frac{1}{2}$ or
$1$ depending on whether the Laplace integral is evaluated at an
interior or a boundary point of the domain.

\begin{figure}[t]
  \begin{center}
   \includegraphics[width=3.5in,angle=-90]{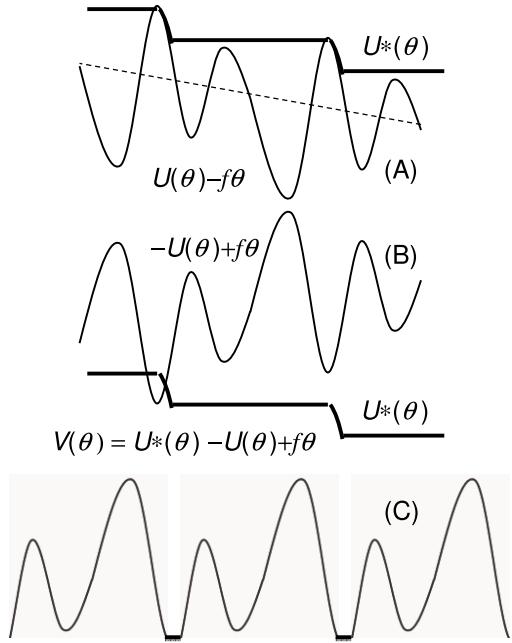}
   \caption{(A) The thin solid line is $U(\theta)-f\theta$,
where $f$ is represented by the slope of the dashed line. The thick
solid line is $U^*(\theta)=\sup_{z\in [\theta,\theta+1)}
\left\{U(z)-fz\right\}$.  When combining $U^*(\theta)$ with
$-U(\theta)+f\theta$, as shown in (B), one obtains $V(\theta)$
given in (C).  $V(\theta)$ is periodic but contains
non-differentiable points.  If $U(\theta)-f\theta$ is monotonically
decreasing, then $U^*(\theta)=U(\theta)-f\theta$ and $V(\theta)=0$.
}
  \label{fig1}
  \end{center}
\end{figure}

    $u(\theta)$ in Eq. (\ref{uss}), therefore,
has the form
\begin{equation}
    u(\theta) = A\ C(\theta,f) \epsilon^{\nu}\
        e^{\frac{V(\theta)}{\epsilon}},
\end{equation}
in which
\begin{equation}
        V(\theta) = U^*(\theta)-U(\theta)+f\theta
\end{equation}
is periodic.  Fig. \ref{fig1} shows one example of how
$U^*(\theta)$ is obtained from $U(\theta)-f\theta$, and $V(\theta)$
is obtained from $U^*(\theta)$.   In general, $V(\theta)$ will have
points of non-differentiability.

    If the periodic $V(\theta)$ is not a
constant, then it reaches its global maximum at a certain
$\theta^*$.  Then in the limit of $\epsilon\rightarrow 0$, the
stationary distribution $u(\theta)\rightarrow \delta
(\theta-\theta^*)$.

    However, if
\[
    f > \max_{\theta\in [0,1]} U'(\theta),
\]
then $U(\theta)-f\theta$ is a monotonically decreasing function of
$\theta$.  In this case, $U^*(\theta)=U(\theta)-f\theta$ and
$V(\theta)\equiv 0$!  Furthermore, $\nu = 1$ and $C(\theta,f) =
1/(f-U'(\theta))$.  Thus in the limit of $\epsilon\rightarrow 0$, we
have
\begin{equation}
    u(\theta) = \left(\int_0^1\frac{dz}{f-U'(z)}\right)^{-1}
            \frac{1}{f-U'(\theta)}.
\end{equation}

    Thus, the stationary distribution $u(\theta)$
reflects the non-uniform velocity on the circle in accordance with
ergodic theory.  The nature of a stable limit cycle being an
attractor, however, is reflected by the constant $\phi(x)$ on the
limit cycle, which has a dynamics on a different time scale when
$\epsilon$ is small.

\subsection{A simple example of diffusion on a circle}

We now give a simple example: A nonlinear dynamics on
a circle $\dot{\theta}=f-\sin(2\pi\theta)$ \cite{strogatz}.
The corresponding Eq. (\ref{1d_s_fpe})
has a $U(\theta)=-1/(2\pi)\cos(2\pi\theta)$, $\theta\in\mathbb{S}[0,1]$. For $f<1$, the deterministic dynamics has a stable fixed
point at $\theta^*$ and unstable fixed point at
$\frac{1}{2}-\theta^*$, where we denote $\theta^*=\frac{1}{2\pi}\arcsin f$, $\theta^*\in\left[0,\frac{1}{4}\right]$.
But for $f>1$, it has no fixed point; instead it has a
limit cycle.  With periodic boundary condition, the
stationary solution to Eq. (\ref{1d_s_fpe}) is
Eq. (\ref{uss}) in the form of
\begin{equation}
      u_{\epsilon}(\theta) = A_{\epsilon}\left(\int_{\theta}^{1+\theta}
    e^{-\frac{1}{\epsilon}\left(\frac{1}{2\pi}
        \cos(2\pi z)+fz\right)}dz\right)
    e^{\frac{1}{\epsilon}\left(\frac{1}{2\pi}
                     \cos(2\pi\theta)+f\theta\right)},
\end{equation}
in which
\[
 A_{\epsilon} = \left[\int_0^1
       \left(\int_{\theta}^{1+\theta}
    e^{-\frac{1}{\epsilon}\left(\frac{1}{2\pi}
        \cos(2\pi z)+fz\right)}dz\right)
    e^{\frac{1}{\epsilon}\left(\frac{1}{2\pi}
                     \cos(2\pi\theta)+f\theta\right)}
                  d\theta\right]^{-1}.
\]

It is easy to show that when $f>1$,
$1/(2\pi)\cos(2\pi z)+fz$ is a monotonically increasing
function of $z$.  Hence applying Laplace's method near
$z=\theta$ one has \cite{bender}
\begin{equation}
   u_0(\theta)
     = \lim_{\epsilon\rightarrow 0} u_{\epsilon}(\theta)
     =  \frac{\sqrt{f^2-1}}{f-\sin(2\pi\theta)},
             \hspace{0.2in} (f>1).
\label{u0theta}
\end{equation}
This is the case with deterministic limit cycle.  According
to the ergodic theory, $u_0(\theta)\propto 1/\dot{\theta}$.

When $f\le 1$, one again applies Laplace's method.
We introduce  $\tilde{\theta}$,
which satisfies
\[
     \frac{1}{2\pi}\cos(2\pi\tilde{\theta})+f\tilde{\theta}
        = -\frac{\cos(2\pi\theta^*)}{2\pi}+f(1/2-\theta^*).
\]
If $\tilde{\theta}\in\left[0,\frac{1}{2}-\theta^*\right]$, then
one has
\begin{equation}
   u_\epsilon(\theta) \approx A_{\epsilon}\exp\left[
         \frac{1}{\epsilon}
         \left\{\begin{array}{ccc}
         \displaystyle  0
                     && 0\le\theta\le\tilde{\theta}
\\[15pt]
         \displaystyle
         \frac{\cos(2\pi\theta)+\cos(2\pi\theta^*)}{2\pi}
           +f\left(\theta+\theta^*-\frac{1}{2}\right)
           && \tilde{\theta}\le\theta\le\frac{1}{2}-\theta^*
\\[15pt]
         \displaystyle  0
                     && \frac{1}{2}-\theta^*\le 1
         \end{array}\right]\right..
\label{fle11}
\end{equation}
If $\tilde{\theta}\in\left[-\frac{1}{2}+\theta^*,0\right]$,
then we denote $\hat{\theta}=1+\tilde{\theta}\in\left[\frac{1}{2}+\theta^*,1\right]$.  Then we have
\begin{equation}
   u_\epsilon(\theta) \approx A_{\epsilon}\exp\left[
         \frac{1}{\epsilon}
         \left\{\begin{array}{ccc}
         \displaystyle
         \frac{\cos(2\pi\theta)+\cos(2\pi\theta^*)}{2\pi}
           +f\left(\theta+\theta^*-\frac{1}{2}\right)
           && 0 \le\theta\le\frac{1}{2}-\theta^*
\\[15pt]
         \displaystyle  0
                     && \frac{1}{2}-\theta^*\le\theta\le\hat{\theta} \\[15pt]
         \displaystyle
         \frac{\cos(2\pi\theta)+\cos(2\pi\theta^*)}{2\pi}
        +f\left(\theta+\theta^*-\frac{3}{2}\right)
                     && \hat{\theta}\le\theta\le 1
                \end{array}\right]\right.
\label{fle12}
\end{equation}
Note that both Eqs. (\ref{fle11}) and (\ref{fle12})
are periodic function of $\theta$ on $[0,1]$.  The
exponents in both are non-negative with a ``flat region''
of zero as $V(\theta)$ illustrated in Fig. \ref{fig1}C.  Furthermore, the maximum of $V(\theta)$ is located at
$\theta^*$, the stable fixed point of the nonlinear dynamics.
Therefore, the normalized $u_0(\theta)=\delta(\theta-\theta^*)$
for $f<1$.

\begin{figure}[t]
  \begin{center}
   \includegraphics[width=5in]{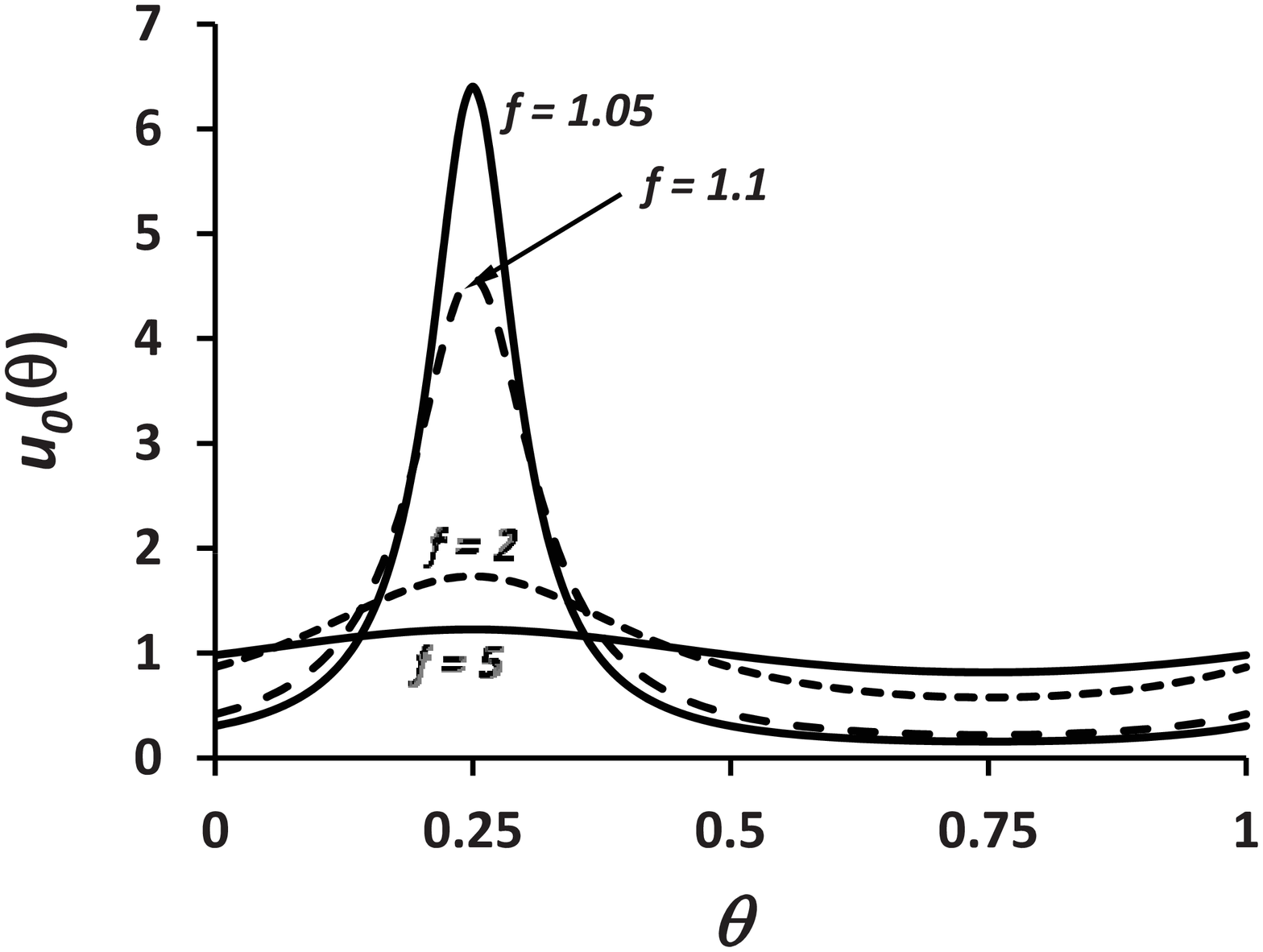}
\caption{The limiting distribution $u_0(\theta)$
according to Eq. (\ref{u0theta}) for nonlinear
dynamics on a circle
$\dot{\theta}$ $=$ $f-\sin(2\pi \theta)$,
with $f=5,2,1.1$ and $1.05$.
With $f\rightarrow 1^+$, it approaches to $\delta(\theta-0.25)$.
For $f\le 1$, the distribution is $\delta(\theta-\theta^*)$
where $\theta^*=1/(2\pi)\arcsin(f)$.
}
  \label{fig2}
  \end{center}
\end{figure}

Note that the limit of Eq. (\ref{u0theta})
\begin{equation}
           \lim_{f\rightarrow 1^+}
                  \frac{\sqrt{f^2-1}}{f-\sin(2\pi\theta)}
              =\delta\left(\theta-\frac{1}{4}\right).
\end{equation}
This simple example has been discussed widely in the
nonlinear dynamic literature on synchronization and
neural networks \cite{strogatz}.  As expected for
the attractor of a nonlinear dynamics with limit
cycle, the law of large numbers for the corresponding
stationary process is not a set of Dirac-delta measures,
but a continuous one.  Fig. \ref{fig2} shows
$u_0(\theta)$ in Eq. (\ref{u0theta}) for several
different values of $f$.

\section{General derivation for high dimensional systems with a Limit Cycle}

The existence of limit cycles is indicative of a system being
far from equilibrium (detailed balance). It has been widely believed that
systems with limit cycle can not have a Lyapunov function.  This is
certainly true according to the strict definition of a Lyapunov
function \cite{perko}. However, in a broader sense, functions with
Lyapunov properties can be constructed for systems with limit cycle.
We shall now consider a multi-dimensional $\vF(\vx)$ and its
corresponding $\phi(\vx)$, as defined above.

    First, we observe that the Lyapunov property of $\phi$
immediately leads to the conclusion that $\phi(\vx)=$ constant
if $\vx\in\Gamma$, where $\Gamma$ is a limit cycle of $\vF$.
To show this, we simply note that
\begin{equation}
    \oint_{\Gamma} \nabla\phi \cdot d\vec{\ell}
     = 0,
\end{equation}
where the integrand
\begin{equation}
     \nabla\phi \cdot d\vec{\ell}
    =\frac{\nabla\phi\cdot\vF\ d\ell}{\|\vF\|}
    \le 0.
\end{equation}
Hence, $\nabla\phi(\vx)=0$, i.e., $\phi(\vx)=$ const, where
$\vx\in\Gamma$. Moreover in any small neighborhood of $\Gamma$,
$\phi(x)$ must be all greater or less than that on $\Gamma$ in the
cases of stable or unstable limit cycles respectively. Here we do
not consider the very complicated case such as strange attractor.

    We now compute $C_0(\vx)$ on $\Gamma$. First of all, according
to Eq. (\ref{eq12}) and $\nabla \phi(\vx)=0$ on $\vx\in\Gamma$, we
have $$\nabla C_0\cdot\vF+C_0\nabla\cdot\vF=0,$$ i.e. $\nabla\cdot
(C_0\vF)=0$. Then we could pick any continuous segment of $\Gamma$,
and consider its $\delta$-thickness neighborhood. When $\delta$
tends to zero, the only fluxes remain are the influx and outflux of
the vector field $C_0\vF$ along $\Gamma$, hence according to Gauss'
theorem, the values of the function $C_0\vF$ at the two ends of the
segment must be the same. Therefore, $\| C_0\vF\|
=C_0\| \vF\|$ must be constant along
$\Gamma$.

Now we restrict the dynamics to $\Gamma$ and introduce an
specific angular variable $\theta$, $\theta\in\mathbb{S}$: $\theta$
is just the arc length starting from some fixed point on $\Gamma$.
In this case, we have $\sum_i (\frac{d\vx_i}{d\theta})^2\equiv 1$.

Then let
\begin{equation}
    \Theta(\theta) = \mathbf{F_i(x(\theta))}
        \left(\frac{d\vx_i}{d\theta}\right)^{-1}
    = \|F\|, \ \ \ i=1,2,\cdots,N,
\end{equation}
where $\vx(\theta)=$ $(\vx_1(\theta),\vx_2(\theta),\cdots, \vx_N(\theta))$
and
\[
    \| F\| =\sqrt{\sum_i (F_i)^2},
\]
such that the differential equation on the limit cycle
$\Gamma$ becomes
\begin{equation}
    \frac{d\theta}{dt}=\Theta(\theta).
\end{equation}
Hence
\begin{equation}
    C_0(\theta) = \frac{A}{\Theta(\theta)},
\label{eq17}
\end{equation}
where $A$ is a normalization constant, whose meaning is very clear:
According to ergodic theory, the stationary probability distribution
of $\theta$ is simply the inverse of the angular velocity.  In fact,
the period of the limit cycle is
\begin{equation}
    T = \int_0^{2\pi}\frac{d\theta}{\Theta(\theta)}.
\end{equation}
This corresponds to the flux on $\Gamma$, i.e, the number of
cycles per unit time, according to Eq. (\ref{ssJ}):
\begin{equation}
    J = A = \left(\int_0^{2\pi}\frac{d\theta}{\Theta(\theta)}
            \right)^{-1}.
\end{equation}
$J$ is also known as the rotation number in nonlinear dynamics.

    We shall note that while $\phi(\vx)$ has the Lyapunov
property, the stationary probability in Eq. (\ref{u_exp}) does not.
The stationary $u_{\epsilon}(\vx)$ can not be a Lyapunov function since in
general it is not a constant on $\Gamma$ due to the contribution
from $C_0(\vx)$ \cite{ao_2006}.

   If one chooses another angular parameter
$\theta'$, then the reciprocal of the velocity
$\Theta(\theta')$ will not be $C_0$. It is straightforward
to modify the analysis presented above.

\subsection{Beyond limit cycle}

The discussion in this section is only heuristic; a more detailed
mathematical analysis remains to be developed.  From the result
above, it seems reasonable that for a high-dimensional nonlinear
ordinary differential equation with vector field $\vF(\vx)$, its
entire center manifold has a constant $\phi(\vx)$, if it exists.
Similarly, $\phi(\vx)$ will be a constant on an invariant torus,
i.e., quasi-periodic motion occurs.  This is easy to illustrate from the
simple example:
\begin{subequations}
\label{torus}
\begin{eqnarray}
    d\theta &=& \Theta\ dt + \sqrt{2\epsilon D_1}\ dB_t^{(1)},
\\
    d\xi &=& \Xi\ dt + \sqrt{2\epsilon D_2}\ dB_t^{(2)},
\end{eqnarray}
\end{subequations}
in which $(\theta,\xi)\in\mathbb{S}^2$.  When $\Theta/\Xi$ is
irrational, the entire $\mathbb{S}^2$ is an invariant torus.
However, the stationary probability for Eq. (\ref{torus}) is
separable in $\theta$ and $\xi$.  Hence according to the
above results on the limit cycle, $\phi(\theta,\xi)$ is
constant on the entire $\mathbb{S}^2$.

\section{Local and global landscapes in the case of multiple attractors and emergent nonequilibrium steady state}

    Nonlinear dynamics on a circle, i.e., Eq. (\ref{1d_s_fpe}),
can only be one of the three types: ($a$) single stable fixed point (attractor), ($b$) multiple stable fixed points (attractors),
and ($c$) oscillation.  Our focus so far has been mainly on ($c$),
and transition from ($a/b$) to ($c$).  However, even within
($a$) and ($b$), there are further distinctions between
gradient systems with $f=0$ and non-gradient system
with $f\neq 0$.  The latter is known as irreversible
diffusion processes \cite{zqq}.  To complete the analysis, we now
consider ($b$). For small $\epsilon$, the dynamics exhibits two different time scales:
intra-attractoral dynamics and inter-attractoral dynamics. The major
questions here are ($i$) the relative stability of these attractors
and ($ii$) the transition rates between different attractors.  Note
the unique feature of dynamics on the circle $\mathbb{S}$, which is
different from one-dimensional $\mathbb{R}$, is the possibility of
non-gradient, i.e., no detailed balance. Stationary, reversible diffusion process on
$\mathbb{R}$ has a single global landscape which simultaneously
provides answers to both ($i$), e.g., Boltzmann's law, and
($ii$) via Kramers' theory \cite{Kramers,gehao_1,gehao_2}.
This is not the case for stationary diffusion process on
$\mathbb{S}$. Although the fundamental theorems by Freidlin and Wentzell have been
developed for quite a long time \cite{freidlin}, their relation to nonequilibrium
thermodynamics is still unknown. The present study, thus,
serves an initiation for this interesting problem.

According to Freidlin and Wentzell \cite{freidlin}, in
the high-dimensional case
as well as in the one-dimensional compact manifold, there are
two types of landscapes: The {\em local landscape}
underlies a Kramers' theory-like analysis for a single
transition from one basin of attraction to another. The
{\em global landscape}, on the other hand, is for the relative stability in nonequilibrium steady state \cite{zqq,gqq}. The
well-known Kramers' rate theory states that \cite{Kramers,gardiner}
the barrier crossing time is exponentially dependent
on the barrier height and nearly exponential distributed \cite{bovier} when the noise strength tends
to zero. Then, putting together all the transition rate
constants computed from the Kramers' theory,
one obtains a discrete-state continuous-time Markovian
chain(In chemistry, this is called discrete chemical
kinetics.).  According to a key theorem in \cite{zqq}, one could then
realize that such a Markov chain is equilibrium
if and only if the local landscapes derived from the
Freidlin-Wentzell local actions in each attractive domain could be
continuously pieced together; The function pieced
together is just the global landscape; this could only be
guaranteed with the detailed balance condition.

Next, for each single domain (or basin) of attraction associated
with a stable fixed point, applying the large deviation theory
of Freidlin and Wentzell, one builds a local landscape
$\phi_i(x)$, $i=1,2,...,N$. And then for each pair of neighbouring attractive domains $\Omega_i$ and
$\Omega_j$, one obtains a pair of local transition rates from
Kramers' theory: the transition rate $k_{ij}$ from $\Omega_i$ to $\Omega_j$ is
proportional to $e^{-V(i,j)/\epsilon}$, where $V(i,j)$ is the lowest
barrier height of $\phi_i(x)$ along the boundary with the attractive
domain $\Omega_j$.

\begin{figure}[t]
  \begin{center}
   \includegraphics[width=4in]{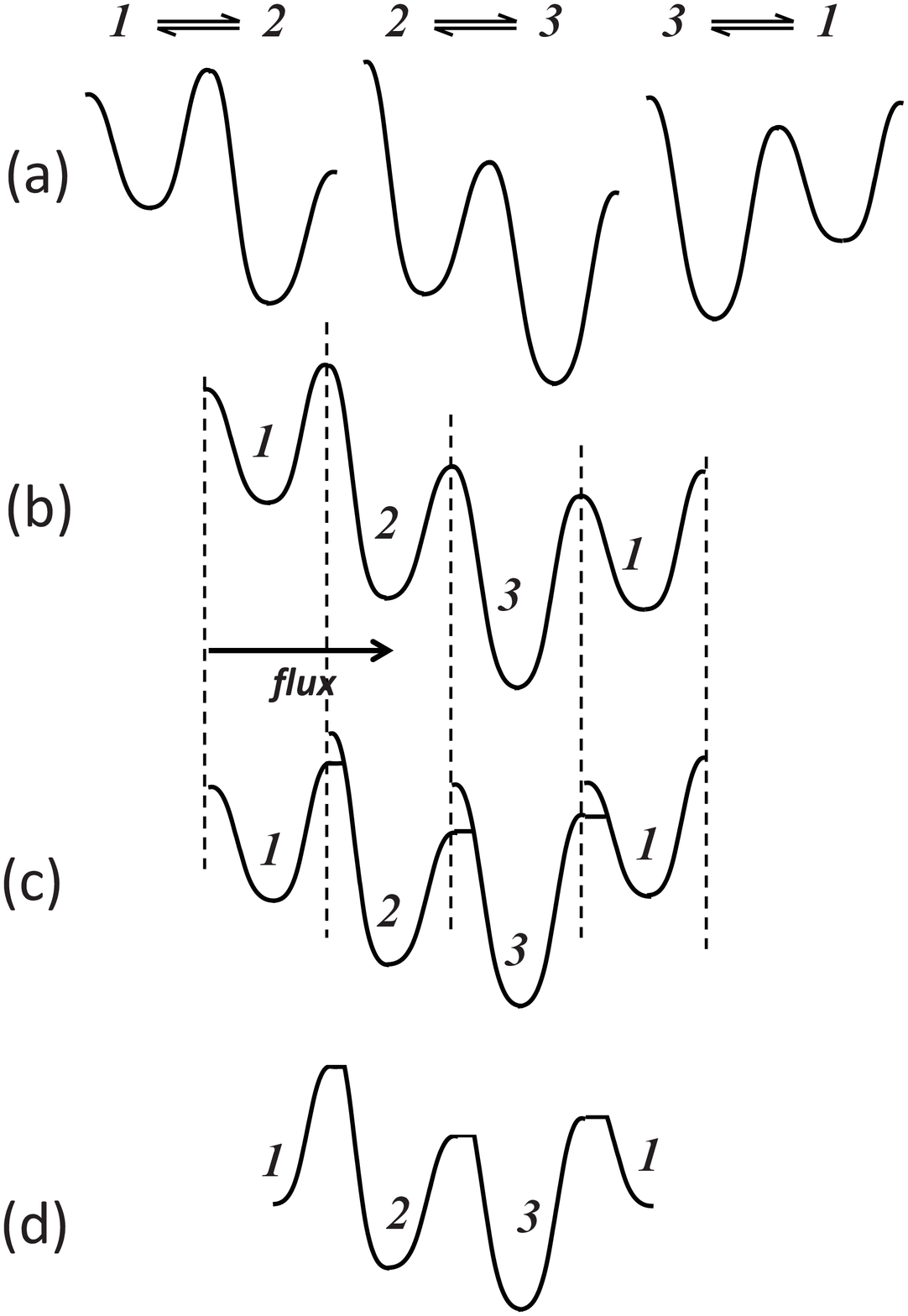}
   \caption{(a) Pairwise local landscapes; (b) A simple ``pasting together'' leads to discontinuous matched case; (c) The global landscape is obtained by a ``$\lambda$-surgery and pasting''
procedure: The surgery lifts the well-2 with respect to
well-1 an amount of $\ln(\pi_1k_{12})/(\pi_2k_{21})$,
which is precisely the free energy difference
$\Delta\mu_{12}$ for
well-2 with respect to well-1 in nonequilibrium steady state \cite{zqq}.
Similarly it lifts the amount of
$\Delta\mu_{23}=\ln(\pi_2k_{23})/(\pi_2k_{32})$ for
well-3 with respect to well-2, and
$\Delta\mu_{31}=\ln(\pi_3k_{31})/(\pi_1k_{13})$ for well-1
with respect to well-3.  Therefore, the total lift is
$\Delta\mu_{12}+\Delta\mu_{23}+\Delta\mu_{31}$ $=$
$\ln(k_{12}k_{23}k_{31})/(k_{21}k_{32}k_{13})$; (d) The final global landscape.
Note that $-V(x)$ in Fig. \ref{fig1}C is just one example of such a global
landscape.  It is a piecewise smooth function with
``flat regions'' at its local maxima.
}
  \label{fig3}
  \end{center}
\end{figure}

This way, we obtain an emergent, discrete-state Markov network
with state space
$\{1,2,...,N\}$ and transition rates $K=\{k_{ij}\}_{N\times N}$
for non-diagonal elements.  The diagonal elements of $K$
are determined by requiring all its rows summed to zero.
A stationary distribution can then be solved, with
$\pi=\{\pi_i\}$ satisfying
$$\pi K=0,$$
Now we are ready for a cricial step: to paste (reshuffle) the local landscapes together
in order to build the global one.
The reshuffle procedure is somewhat subtle. There is an illustrative
example in \cite{freidlin} (also see Fig. \ref{fig3}). Here we
give a simple demonstration in terms of a $3$-state Markov chain.

In this case, we have
$$\pi_1=\frac{k_{23}k_{31}+k_{31}k_{21}+k_{21}k_{32}}{\mathcal{D}},$$
$$\pi_2=\frac{k_{31}k_{12}+k_{12}k_{32}+k_{32}k_{13}}{\mathcal{D}},$$
$$\pi_3=\frac{k_{12}k_{23}+k_{23}k_{13}+k_{13}k_{21}}{\mathcal{D}},$$
in which the denominator $\mathcal{D}$ is determined by $\pi_1+\pi_2+\pi_3=1$.

When $\epsilon$ tends to zero, let
$$W_1=-\lim_{\epsilon\rightarrow\infty}\epsilon\log\pi_1=\min\{V(2,3)+V(3,1), V(3,1)+V(2,1), V(2,1)+V(3,2)\},$$
$$W_2=-\lim_{\epsilon\rightarrow\infty}\epsilon\log\pi_2=\min\{V(3,1)+V(1,2), V(1,2)+V(3,2), V(3,2)+V(1,3)\},$$
$$W_3=-\lim_{\epsilon\rightarrow\infty}\epsilon\log\pi_3=\min\{V(1,2)+V(2,3), V(2,3)+V(1,3), V(1,3)+V(2,1)\}.$$
So the reshuffle rule is as follows
$$W(x)=\min\{W_1+V(1,x), W_2+V(2,x), W_3+V(3,x)\},$$
where $V(i,x)$ means the minimum of the Freidlin-Wentzell action
along the path between the $i$-th attractive domain and the position
$x$. The intuitive understanding is that starting from each
attractor, and compare their probabilities for arriving at $x$.

In the case of three states, for instance, if $x$ is in the first
attractive domain, then $V(1,x)=\phi_1(x)$, $V(2,x)=V(2,1)$ if the
backward trajectory starting from $x$ would arrive at the boundary
between the first and second attractive domains, otherwise
$V(2,x)=V(2,1)+\phi_1(x)$;  $V(3,x)=V(3,1)$ if the backward
trajectory starting from $x$ would arrive at the boundary between
the first and third attractive domains, otherwise
$V(3,x)=V(3,1)+\phi_1(x)$.

Hence, $W(x)$ is continuous, and the global invariant distribution
\begin{equation}
  u(x)\propto e^{-(W(x)-\min_i\{W_i\})/\epsilon}.
\end{equation}
So $W(x)-\min_i\{W_i\}$ is the global landscape, which also satisfies
the Lyapunov property.

Furthermore, we know that the emergent Markovian chain is in
equilibrium, if and only if $k_{12}k_{23}k_{31}=k_{21}k_{32}k_{13}$  \cite{zqq},
which means $V(1,2)+V(2,3)+V(3,1)=V(2,1)+V(3,2)+V(1,3)$, i.e.
$W_2-W_1=V(1,2)-V(2,1)$, $W_3-W_2=V(2,3)-V(3,2)$ and
$W_1-W_3=V(3,1)-V(1,3)$. Hence the local landscape $\phi_i(x)$ would
be continuously connected at the boundaries in this case.

The above result is a generalization of the celebrated work of Kramers \cite{Kramers}.  In Kramers' theory, the underlined nonlinear diffusion process is the atomic dynamics along the reaction coordinate,
while the emergent discrete dynamics is exactly the discrete chemical kinetics.  For systems with detailed balance,
Kramers' rate constants are consistent with Boltzmann's law for
conformational probabilities.  However, when detailed balance
are not satisfied, we have clearly demonstrated here an
essential difference between local and global landscapes: The
former is related to individual state-to-state transition,
while the latter is associated with a systems' long-time
dynamics. Their disagreement is the origin of nonequilibrium
steady states \cite{zqq,gqq,qian_jpc}.

\section{Conclusions}

    Multi-dimensional diffusion processes and Markov jump
processes with chemical master equations are two
mathematical models for studying mesoscopic, nonequilibrium
physical and biochemical dynamics with multiscale phenomena
and emergent organizations \cite{qian_iop}. In
the past, our understandings of dynamics in terms of its
molecular constituents have been mainly derived from theories
of macroscopic, deterministic dynamics in the thermodynamic
limit \cite{kurtz} or statistical mechanics of
closed systems which are necessarily equilibrium.
Much to be learned from the two types of stochastic models
for the mesoscopic dynamics in open systems, especially
the relationship between their asymptotic dynamics
and emergent nonequilibirum steady state.  Even for the
simplest case of one-dimensional circle, there were
important questions to be addressed and answered.
In the present work, insights have been gained from
applying methods of singular perturbation and the
theory of large deviations. It has been shown that the
intuitive notion of a landscape can be further
secured by applying the mathematical theories.


Combining the insights with a wide range of existing
applied mathematical techniques (see several
reviews \cite{ludwig_75,schuss_80,omalley_08,matkowsky_2}),
the study illustrated here
can be further taken into several directions.
Strengthening the tie \cite{roy_93} between the abstract
theory of large deviations \cite{freidlin} and more applied
singular perturbation techniques \cite{bender,grasman,kuske}
will yield further understandings for stochastic
nonlinear dynamics (SND).  In particular,
the large deviation behavior of Delbr\"{u}ck-Gillespie process
\cite{qian_iop} is still poorly understood.
The results from many previous workers synthesized in the
present work also provides a glimpse of how to develop an alternative structural stability theory for nonlinear
dynamical systems, as called by E.C.
Zeeman many years ago \cite{zeeman}.  One naturally wonders what
$\phi$ and $C_0$ will be for a chemical reaction system that
possesses a chaotic attractor \cite{epstein}.
Is there any regularity in the asymptotic behavior of the
invariant measure for such systems \cite{kuske,lsyoung,xie_qian,weinan}?
These are hard problems; but they are no longer impossible to conceive.

\vskip 1.5cm

    HQ thanks Ping Ao, Bernard Deconinck, Gang Hu, Rachel Kuske,
Robert O'Malley and Jin Wang for many fruitful discussions.
HG acknowledges support by NSFC 10901040,
specialized Research Fund for the Doctoral Program of Higher
Education (New Teachers) 20090071120003 and the Foundation for the
Author of National Excellent Doctoral Dissertation of China (no 201119).

\end{document}